\documentclass[11pt,a4paper]{article}
\usepackage{amssymb} \usepackage{amsmath} \usepackage{graphicx}
\usepackage{epsfig,latexsym,color,xcolor}
\usepackage{ulem}
\usepackage{amstext}    
\usepackage{array} 
\usepackage{slashed}
\begin{document}
\def\Giulia{\bf\color{red}}
\def\bef{\begin{figure}}
\def\eef{\end{figure}}
\newcommand{\ans}{ansatz }
\newcommand{\be}[1]{\begin{equation}\label{#1}}
\newcommand{\beq}{\begin{equation}}
\newcommand{\ee}{\end{equation}}
\newcommand{\beqn}[1]{\begin{eqnarray}\label{#1}}
\newcommand{\eeqn}{\end{eqnarray}}
\newcommand{\bd}{\begin{displaymath}}
\newcommand{\ed}{\end{displaymath}}
\newcommand{\mat}[4]{\left(\begin{array}{cc}{#1}&{#2}\\{#3}&{#4}
\end{array}\right)}
\newcommand{\matr}[9]{\left(\begin{array}{ccc}{#1}&{#2}&{#3}\\
{#4}&{#5}&{#6}\\{#7}&{#8}&{#9}\end{array}\right)}
\newcommand{\matrr}[6]{\left(\begin{array}{cc}{#1}&{#2}\\
{#3}&{#4}\\{#5}&{#6}\end{array}\right)}
\newcommand{\cvb}[3]{#1^{#2}_{#3}}
\def\lsim{\raise0.3ex\hbox{$\;<$\kern-0.75em\raise-1.1ex
e\hbox{$\sim\;$}}}
\def\gsim{\raise0.3ex\hbox{$\;>$\kern-0.75em\raise-1.1ex
\hbox{$\sim\;$}}}
\def\abs#1{\left| #1\right|}
\def\simlt{\mathrel{\lower2.5pt\vbox{\lineskip=0pt\baselineskip=0pt
           \hbox{$<$}\hbox{$\sim$}}}}
\def\simgt{\mathrel{\lower2.5pt\vbox{\lineskip=0pt\baselineskip=0pt
           \hbox{$>$}\hbox{$\sim$}}}}
\def\unity{{\hbox{1\kern-.8mm l}}}
\newcommand{\eps}{\varepsilon}
\def\ep{\epsilon}
\def\ga{\gamma}
\def\Ga{\Gamma}
\def\om{\omega}
\def\omp{{\omega^\prime}}
\def\Om{\Omega}
\def\la{\lambda}
\def\La{\Lambda}
\def\al{\alpha}
\def\beq{\begin{equation}}
\def\eeq{\end{equation}}
\newcommand{\ov}{\overline}
\renewcommand{\to}{\rightarrow}
\renewcommand{\vec}[1]{\mathbf{#1}}
\newcommand{\vect}[1]{\mbox{\boldmath$#1$}}
\def\tm{{\widetilde{m}}}
\def\mcirc{{\stackrel{o}{m}}}
\newcommand{\Dm}{\Delta m}
\newcommand{\dm}{\varepsilon}
\newcommand{\tanb}{\tan\beta}
\newcommand{\nbar}{\tilde{n}}
\newcommand\PM[1]{\begin{pmatrix}#1\end{pmatrix}}
\newcommand{\up}{\uparrow}
\newcommand{\down}{\downarrow}
\newcommand{\refs}[2]{eqs.~(\ref{#1})-(\ref{#2})}
\def\omE{\omega_{\rm Ter}}
\newcommand{\eqn}[1]{eq.~(\ref{#1})}
%

\newcommand{\DSUSY}{{SUSY \hspace{-9.4pt} \slash}\;}
\newcommand{\DCP}{{CP \hspace{-7.4pt} \slash}\;}
\newcommand{\mc}{\mathcal}
\newcommand{\gr}{\mathbf}
\renewcommand{\to}{\rightarrow}
\newcommand{\gtc}{\mathfrak}
\newcommand{\wh}{\widehat}
\newcommand{\br}{\langle}
\newcommand{\kt}{\rangle}


\def\lsim{\mathrel{\mathop  {\hbox{\lower0.5ex\hbox{$\sim$}
\kern-0.8em\lower-0.7ex\hbox{$<$}}}}}
\def\gsim{\mathrel{\mathop  {\hbox{\lower0.5ex\hbox{$\sim$}
\kern-0.8em\lower-0.7ex\hbox{$>$}}}}}

\def\nn{\\  \nonumber}
\def\de{\partial}
\def\brf{{\mathbf f}}
\def\bbf{\bar{\bf f}}
\def\bF{{\bf F}}
\def\bbF{\bar{\bf F}}
\def\bA{{\mathbf A}}
\def\bB{{\mathbf B}}
\def\bG{{\mathbf G}}
\def\bI{{\mathbf I}}
\def\bM{{\mathbf M}}
\def\bY{{\mathbf Y}}
\def\bX{{\mathbf X}}
\def\bS{{\mathbf S}}
\def\bb{{\mathbf b}}
\def\bh{{\mathbf h}}
\def\bg{{\mathbf g}}
\def\bla{{\mathbf \la}}
\def\bmu{\mathbf m }
\def\by{{\mathbf y}}
\def\bmu{\mbox{\boldmath $\mu$} }
\def\bsig{\mbox{\boldmath $\sigma$} }
\def\bunity{{\mathbf 1}}
\def\cA{{\cal A}}
\def\cB{{\cal B}}
\def\cC{{\cal C}}
\def\cD{{\cal D}}
\def\cF{{\cal F}}
\def\cG{{\cal G}}
\def\cH{{\cal H}}
\def\cI{{\cal I}}
\def\cL{{\cal L}}
\def\cN{{\cal N}}
\def\cM{{\cal M}}
\def\cO{{\cal O}}
\def\cR{{\cal R}}
\def\cS{{\cal S}}
\def\cT{{\cal T}}
\def\eV{{\rm eV}}
%




\large
 \begin{center}
 {\Large \bf Dark Matter from Starobinsky Supergravity }
 \end{center}

 \vspace{0.1cm}

 \vspace{0.1cm}
 \begin{center}
{\large Andrea Addazi}\footnote{E-mail: \,  andrea.addazi@infn.lngs.it} \\
{\it \it Dipartimento di Fisica,
 Universit\`a di L'Aquila, 67010 Coppito, AQ 
LNGS, Laboratori Nazionali del Gran Sasso, 67010 Assergi AQ, Italy}
\end{center}

  \begin{center}
{\large Maxim Yu. Khlopov}ì
\\
{\it Centre for Cosmoparticle Physics Cosmion;
National Research Nuclear University MEPHI (Moscow Engineering Physics Institute), Kashirskoe Sh., 31, Moscow 115409, Russia;
APC laboratory 10, rue Alice Domon et L\'eonie Duquet 75205 Paris Cedex 13, France}
\end{center}

\vspace{1cm}
\begin{abstract}
\large

We review our recent results on dark matter from Starobinsky supergravity.
 In this context, a natural candidate for Cold Dark Matter is the gravitino.
On the other hand, assuming the supersymmetry broken at
scales much higher than the electroweak scale, 
gravitinos are super heavy particles. 
In this case, they may be  non-thermally produced during inflation, 
in turn originated by the scalaron field with Starobinsky's potential.
 Assuming gravitinos as Lightest supersymmetric particles (LSSP),
  the non-thermal production naturally accounts for the 
 right amount of cold dark matter. Metastability of the gravitino LSSP leads to observable effects of their decay, putting constraints on the corresponding Unstable or Decaying Dark Matters scenarios.
In this model, the gravitino mass is controlled by the inflaton field
and it runs with it. This implies that a continuous spectrum 
of superheavy gravitinos is produced during the 
slow-roll epoch. 
Implications in phenomenology, model building in 
GUT scenarios, intersecting D-branes models
and instantons in string theories 
are discussed.

\end{abstract}

\baselineskip = 20pt


\section{Introduction}
Direct searches for Weakly Interacting Massive Particles supersymmetric Dark Matter candidates do not give positive results as well as TeV-scale supersymmetry was not found at the LHC.
It may move the supersymmetry scale to much higher energies.
On the other hand, the Starobinsky $R+\zeta R^{2}$ model \cite{S1} shows a substantially good agreement with Recent Planck data \cite{Ade:2015lrj}. 
In particular, 
Starobinsky's model is conformally equivalent to 
a scalar-tensor theory and the scalar field is a 
slow-rolling inflaton.  
.
This may motivate a supergravity
reformulation of the old Starobinsky model, 
assuming supersymmetry spontaneously broken at very high scales. 

A consistent embedding of the old Starobinsky model in supergravity 
is not so easy as naively expectable. 
For instance, it was realized the 
the first Starobinsky supergravity model
proposed in Refs.\cite{9,11} entails
a tachyonic instability of the Goldstino field 
for large values of the inflaton field. 
Recently, these issues were revisited 
 in Refs.\cite{5,6}
and in Refs.\cite{18a,18b,18c,18d,18e,21,22,Ferrara:2013rsa,Ferrara:2013pla,Ferrara:2014cca,Ferrara:2015ela,Ozkan:2014cua,Ferrara:2014kva,Antoniadis:2014oya,Dudas:2015eha,Ferrara:2015gta,Ferrara:2016buf}:
Starobinsky supergravity was reformulated in 
frameworks of 
non-linear Volkov-Akulov supersymmetry and no-scale invariance. 
This new class of models allows to reformulate 
a consistent $R+\zeta R^{2}$ supergravity 
 without any pathologically unstable moduli fields. 
 Recently, the consistency of 
Starobinsky supergravity 
with Null and Weak energy 
conditions was also discussed 
in Ref.\cite{Addazi:2017rkc}. In Ref.\cite{Addazi:2017rkc}, we have also demonstrated that 
the Strong Energy condition
is violated in a large region 
of parameter spaces, 
compatible with inflation. 
 
However,  new Starobinsky supergravity models 
do not only consistently contain old Starobinsky inflation
but they may provide a new candidate of dark matter. 
In particular, these models must predict the 
presence of gravitinos, which, 
if in turn assumed as the lightest supersymmetric particle, 
 may provide 
a natural candidate for cold dark matter. 
Of course, gravitino mass is highly 
dependent by supersimmetry breaking scale. 
So that Recent LHC constraints on TeV-ish SUSY may 
motivate the analysis of superheavy gravitinos.
On the other hand, this opens new issues regarding 
the production of gravitons: may they account for
the right amount of cold dark matter?
In Ref.\cite{Addazi:2016bus}, we have 
provided the analysis regarding this
issue. 

Here we will review our recent results 
obtained in Ref.\cite{Addazi:2016bus}. 
We have studied  $R+\zeta R^{2}$ supergravity 
 with local supersymmetry broken 
at scales higher than the inflaton reheating. 
We will show how super-heavy gravitinos are non-thermally produced during Starobinsky's inflation.
In this mechanism,  
 the right Cold Dark Matter abundance, without any WIMP-like thermal miracle. 
On the other hand, assuming the gravitino mass 
heavier than then the inflaton mass, 
the suppression of the gravitino thermal production 
was shown in 
Ref. \cite{Rychkov:2007uq}.

\section{Gravitino mass in Starobinsky supergravity}

The formulation of the 
Starobinsky supergravity is based on the following 
 Lagrangian 
\cite{Ferrara:2013rsa,Ferrara:2013pla,Ferrara:2014cca,Ferrara:2015ela,Antoniadis:2014oya}:
\be{L}
\mathcal{L}=-[- L\mathcal{V}_{R}+L\Phi(z,\bar{z})]_{D}+\zeta[\mathcal{W}_{\alpha}(\mathcal{V}_{R})\mathcal{W}^{\alpha}(\mathcal{V}_{R})]
\ee
$$\mathcal{V}_{R}=\log \frac{L}{\mathcal{S}\mathcal{\bar{S}}}$$
where  the standard Einstein-Hillbert action is recovered by the first term, 
 the higher derivative term $R^{2}$ term is originated from the kinetic term of the (real) superfield $\mathcal{V}_{R}$,
 $\mathcal{S}$ is the so dub {\it compensator field} of minimal supergravity, 
 $\Phi(z,\bar{z})$ is the K\"ahler potential of the $z,\bar{z}$ fields and 
$L$ is the linear multiplet. 

For a discussion of the gravitino mass 
from supergravity, it 
is convenient to consider the 
 off-shell formulation of the minimal Starobinsky lagrangian during inflation.
 Its K\"ahler potential reads as
\be{K}
\mathcal{K}=-3\log [\mathcal{T}+\mathcal{\bar{T}}-\Phi(z,\bar{z})],\,\,\,\mathcal{W}_{I}\rightarrow 0\,.
\ee
The gravitino mass is directly controlled by the 
K\"ahler potential and the superpotential as
\be{corr}
m_{\tilde{G}}=e^{\mathcal{K}/2}\frac{\mathcal{W}}{M_{Pl}^{2}}=e^{-\sqrt{\frac{3}{2}}\phi} \frac{\mathcal{W}}{M_{Pl}^{2}}
\ee
Let us remark that Eq.(\ref{corr}) implies a direct connection among the 
gravitino mass and the inflaton field. 
In other words, gravitino mass is a functional of the 
inflaton field and it runs with it. 
This will turn out to imply important
predictions in gravitino mass spectrum. 
Let us also note that
$$\mathcal{W}_{I}=0 \rightarrow m_{\tilde{G}}=0\,.$$
Naturally, the gravitino is massless only in the 
supersymmetry and R-symmetry preserving 
phase. As a consequence, the fact 
that the gravitino mass depends on the 
inflaton field is not relevant in this case. 

In our model, as mentioned above, 
we shall assume that supersymmetry is 
{\it  spontaneously broken at scales higher
than the inflation reheating}.
This means that during inflation, 
 the superpotential $\mathcal{W}$ is 
a constant $\mathcal{W}_{0}>0$.
So that, the relation among the gravitino 
mass and the inflaton field is not
more trivial. We will see in the next
sections that this will imply 
that {\it a continuos spectrum of gravitinos  
will be produced during the inflation}.


On the other hand, 
ss mentioned above,  the condition $\mathcal{W}_{I}\rightarrow 0$
during inflation provides a possible way-out to the moduli problems. 
In other words, 
the superpotential may 
roll down to zero before the inflation epoch, 
without causing any back-reactions
to the slow-roll dynamics. 
As remarked above. 
 $\mathcal{W}_{I}=0$ corresponds to a vacuum 
state which is invariant under the R-symmetry and SUSY.
So that, it seems that these symmetries 
may act as a sort of projection 
from the full general case to 
the ones providing a successful
inflation. 
However, 
the condition $\mathcal{W}_{I}=0$ cannot be compatible 
with our dark matter model, because implying massless gravitinos 
during inflation.
So that, we suggest that $U_{R}(1)$ and SUSY 
are spontaneously broken before (or at least during)
the slow-roll epoch. 
On the other hand, after the inflation epoch, 
the rapid rolling down of the superpotential is 
already assumed. 
Under these hypothesis, 
the superpotential may 
only be set to a constant non-zero value
$\rightarrow \mathcal{W}_{0}={\rm  const}\neq 0$. 
To show that this condition does not destabilize 
the model is straightforward. 
For instance, it will imply that 
 the $\mathcal{G}$-term will get an extra constant term:
\begin{equation}
\label{GKK}
\Delta\mathcal{G}=\log \mathcal{W}_{0}+\log \bar{\mathcal{W}}_{0}={\rm const}
\end{equation}
which in turns provides a constant term for the $V_{F,D}$-terms:
\begin{equation}
\label{VF}
\Delta V_{F}=-3 \mathcal{W}_{0}\bar{\mathcal{W}}_{0}
\end{equation}
$$2\zeta\Delta V_{D}=-12\,.$$

So that, the spontaneous symmetry breaking of $U_{R}(1)$
and SUSY during inflation only implies 
that 
 the inflaton potential 
is shifted by a constant, 
which may be reabsorbed in the 
cosmological term. 
But what is important is that 
this demonstrates that 
the spontaneous symmetry breaking 
of $U_{R}(1)$ and SUSY cannot 
destabilize the moduli fields, 
i.e. it does not 
contribute with new extra dangerous interactions term 
in $\mathcal{W}_{I}$. 
For instance, the R-symmetry implemented in the action
has fixed the structure of the potential 
under the condition on $\mathcal{W}_{I}$. 
One can see that the only effect of a $\mathcal{W}_{0}={\rm  const}\neq 0$
during the inflation is the shift of the Starobinsky's potential of a constant factor
and the $z_{I}$ fields remain stabilized.

\section{Non-thermal production of Gravitinos during the slow-roll}

\begin{figure}[htb] \label{GMPLB}
\begin{center}
\caption{Gravitino mass function of Starobinsky inflaton. 
In the x-axis, the inflaton field is conveniently normalized in Planck units, 
while in the y-axis the gravitino mass function is normalized with respect of the average gravitinos mass
$\langle m_{\tilde{G}} \rangle$ (in $log_{10}$ scale in the $y$-axis). 
In particular, the oscillating epoch effectively starts at $\phi\//M_{P}\simeq 1$. 
On the other hand, the slow-roll effectively starts at $\phi/M_{P}\simeq 6$. 
 $\Delta\phi/M_{P}\sim 1\div 6$ is the gravitino production epoch.
So that, a continuos spectrum of super-heavy gravitinos is produced. }
\includegraphics[scale=0.08]{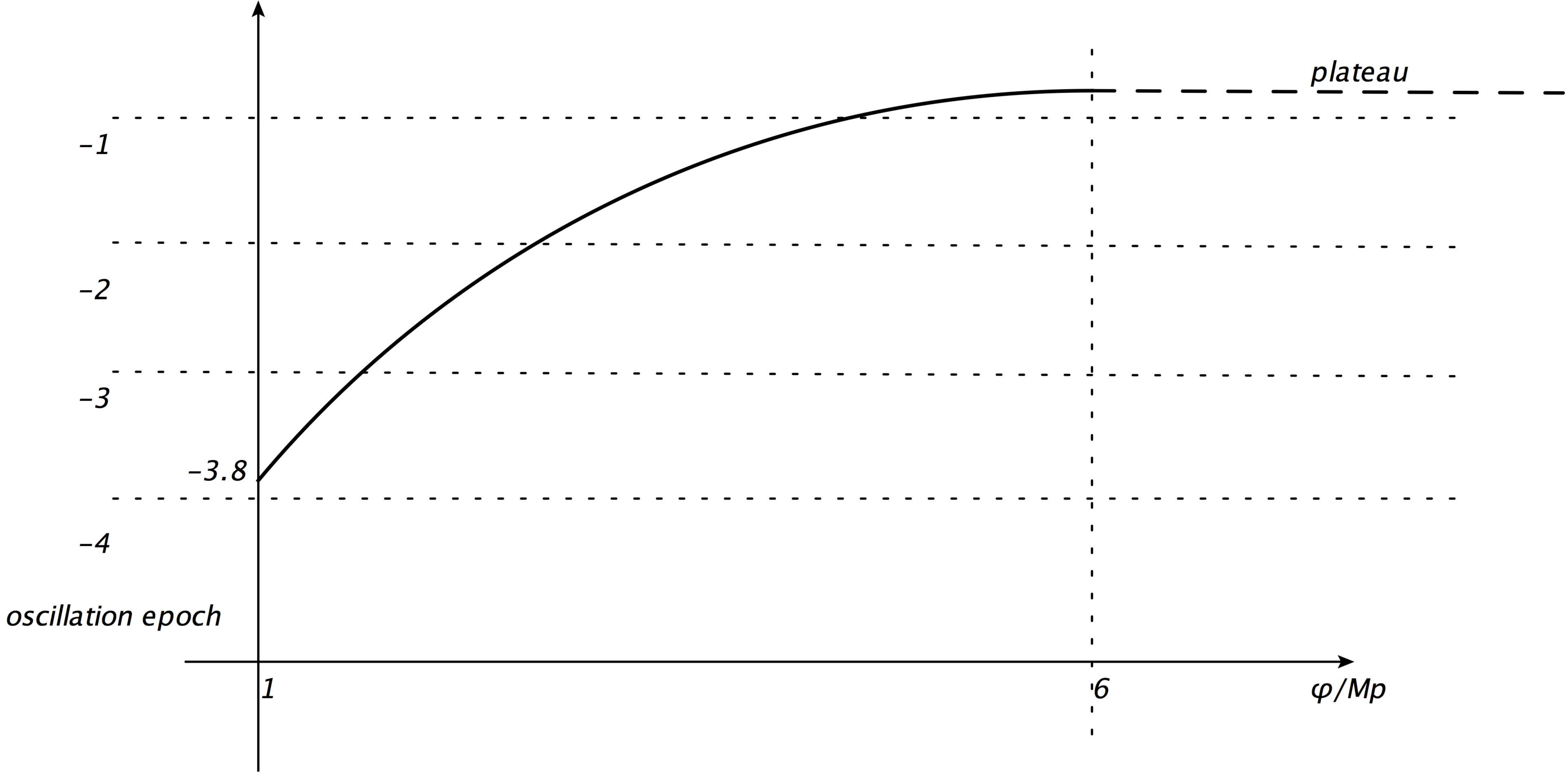}  
\end{center}
\vspace{-1mm}
\end{figure}

In Ref. \cite{Addazi:2016bus}, we have calculated 
the the production rate of gravitinos 
during inflation. 

One can estimate the energy density of the gravitinos produced during inflation
as 
\begin{equation}
\label{estimation}
\rho_{\tilde{G}}(\eta_{e})=\langle m_{\tilde{G}}\rangle n_{\tilde{G}}(\eta_{e})=\langle m_{\tilde{G}}\rangle H_{e}^{3}\left(\frac{1}{a(\eta_{e})} \right)\mathcal{P}
\end{equation}
where $\eta$ is the time-like cosmological time variable, $n_{\tilde{G}}$ is the number density of gravitinos, 
$H_{e}$ is the Hubble rate at the and of the slow-roll epoch time $t_{e}$;
where $\mathcal{P}$
is the power of emission of gravitinos 
from the expanding background
which can be calculated from a
Bogoliubov transformation of 
creation and destruction operators 
associated to the gravitino field
in the expanding FRW background;
where $\langle m_{\tilde{G}}\rangle$
is the average mass of gravitinos produced 
during the slow-roll, which is 
\be{mass}
\langle m_{\tilde{G}}\rangle \simeq \langle e^{-\sqrt{\frac{3}{2}}\phi} \rangle_{\Delta N} \frac{\mathcal{W}_{0}}{M_{Pl}^{2}}\simeq 0.15\frac{\mathcal{W}_{0}}{M_{Pl}}
\ee
considering the inflationary plateau 
has a width of $\Delta \phi\simeq 5M_{Pl}$, i.e $\Delta N=\log a_{f}/a_{i}\simeq 60$ e-folds.
In first approximation, 
one may set in Eq.(\ref{mass}) 
$$\langle \phi \rangle\simeq \Delta \phi/2$$
The mass spectrum is shown in Fig.1, 
as a  function of the cosmological time. 

One can estimate the relation among the gravitino energy density 
normalized over the radiation density. 
It reads as 
\begin{equation}
\label{gravitino}
\frac{\rho_{\tilde{G}}(t_{0})}{\rho_{R}(t_{0})}=\frac{\rho_{\tilde{G}}(t_{R_{e}})}{\rho_{R}(t_{R_{e}})}\left(\frac{T_{R}}{T_{e}} \right)
\end{equation}
where 
$\rho_{\tilde{G}}(t_{Re})/\rho_{R}(t_{Re})$
is the after-Reheating epoch ratio
among gravitinos and radition and where
 $t_{0}$ is the present cosmological time.
 
$\rho_{\tilde{G}}(t_{Re})/\rho_{R}(t_{Re})$
during the reheating epoch -inflaton decays to SM particles-
is estimated as
\begin{equation}
\label{XR}
\frac{\rho_{\tilde{G}}(t_{Re})}{\rho_{R}(t_{Re})}\simeq \frac{8\pi}{3}\left(\frac{\rho_{\tilde{G}}(t_{e})}{M_{Pl}^{2}H^{2}(t_{e})} \right)
\end{equation}
Let us remind that the inflaton mass sets the characteristic scale for the Hubble constant calculated in $t_{e}$:
$H^{2}(t_{e})\sim m_{\phi}^{2}$ and $\rho(t_{e})\sim m_{\phi}^{2}M_{Pl}^{2}$.
This implies 
\begin{equation}
\label{begine}
\Omega_{\tilde{G}}h^{2}\sim 10^{17}\, \left( \frac{T_{Rh}}{10^{9}\, \rm GeV}\right)\left(\frac{\rho_{\tilde{G}}(t_{e})}{\rho_{c}(t_{e})} \right)
\end{equation}
where $\rho_{c}(t_{e})=3H(t_{e})^{2}M_{Pl}^{2}/8\pi$ is the critical energy density during $t_{e}$.
Finally, Eq.(\ref{begine}) can be rewritten as
\begin{equation}
\label{rewe}
\Omega_{\tilde{G}}h^{2}\simeq \Omega_{R}h^{2}\left(\frac{T_{Rh}}{T_{0}} \right)\frac{8\pi}{3}\left( \frac{\langle m_{\tilde{G}}\rangle }{M_{Pl}}\right)\frac{n_{\tilde{G}}(t_{e})}{M_{Pl}H^{2}(t_{e})}
\end{equation}
Eq.(\ref{rewe}) is very useful: it 
relates the gravitino abundance with the gravitino mass, the inflaton mass
and the reheating termperature. 
The inflaton mass is of the order of 
 $m_{\phi} \simeq 10^{13}\, \rm GeV$ or so.
 On  the other hand, the reheating temperature is
$T_{Rh}/T_{0}\simeq 4.2\times 10^{14}$.
These parameters are fixed for a successful 
inflation and reheating. 
So that, the correct abundance of dark matter
is obtained 
for a gravitino mass of
$\langle m_{\tilde{G}}\rangle \simeq (10^{-2}\div 1)\times m_{\phi}$
$\simeq 10^{11}\div 10^{13}\, \rm GeV$,
in turn 
constraining $W_{0}$ in Eq.(\ref{mass}). 
This means that the SUSY symmetry breaking scale
is expected to be around the GUT scale $10^{15\div 16}\, \rm GeV$. 
This certainly leads to other indirect implications in particle physics beyond the standard model.
In fact, if supersymmetry must be broken around the GUT scale,
it does not be helpful for couplings unification in 
GUT scenarios like $SU(5)$ and $SO(10)$.
As a consequence our model seems to motivate 
non-supersymmetric GUT scenarios in which 
the couplings unification is reobtained 
adding extra non-minimal multiplets
  (See for example Ref.\cite{Bertolini:2009es}
  for a revival of non-supersymmetric $SO(10)$ models
  by introducing higher multiplets and considering 
  RG corrections beyond the tree-level relations.).
On the other hand, generically, these multiplets must be added
in order to obtain a realistic spectrum of SM Yukawa 
and neutrino mass matrix \cite{Bajc:2005zf}. 
An alternative paradigm to the unification 
is provided by intersecting D-branes models or quiver string theories. 
In this case, starting from $\mathcal{N}=1$ supersymmetry,
it can be broken 
at the GUT scale (here only taken as conventional) without destabilizing the 
construction, i.e. tachyons in D-brane worldsheets are avoided. 
This certainly seems to be a more promising class of models 
with respect to other attempts to construct intersecting D-brane models 
without supersymmetry, which in general are expected to be 
plagued by tachyons. 
Let us also note that in 
intersecting D-brane models, 
supersymmetry 
may be dynamically broken by the
Euclidean D-brane instantons
\cite{Bianchi:2009bg}.

\section{Phenomenology}

Certainly, Superheavy gravitino dark matter
cannot be searched by direct detection experiments 
or in TeV-scale collider physics. 
However, we will comment how superheavy gravitinos 
may be detected in very high energy indirect
detection experiments, i.e. high energy cosmic rays observations. 

The spontaneously symmetry breaking of 
the gauged $U_{R}(1)$ parity 
may allow to new effective operators destabilizing the 
gravitino and opening new decay channels to 
Standard Model particles. 
The new effective operators which may be 
generated are dependent
by the details of the R-symmetry breaking. 
Of course, in realistic models, such operators
must be very suppressed. Otherwise, the 
gravitino cannot be a good (meta)stable candidate 
for dark matter. 

For example, $U(1)_{R}$ may be spontaneously 
broken by the a scalar singlet field $s$
contained in a supersymmetric chiral field $\mathcal{S}$.
Supposing that $R(\mathcal{S}^{n})=-R(L)$, 
being $R$ the charge operator of $U(1)_{R}$, 
one may introduce effective superpotentials like 
\be{sHL}
\mathcal{W}_{sHL}=\frac{1}{M^{n-1}}S^{n}HL
\ee
where $M$ is an effective suppression scale
generated by UV completion of the model.
This generates an effective operator
$$O_{sHL}=\frac{1}{M^{n-1}}\phi_{S}^{n}\tilde{h}l_{L}$$
where $\tilde{h}$ is the Higgsino field
-it mixes the Higgsino field with neutrinos. 
On the other hand, one can always 
introduce the operator
\be{sg}
L_{\tilde{G}VV}=-\frac{i}{8M_{Pl}}\bar{\psi}_{\mu}[\gamma^{\nu},\gamma^{\rho}]\gamma^{\mu}\lambda F_{\nu\rho}
\ee
coupling the gravitino with $W^{\pm},Z,\gamma$. 
Neutral gauginos mix with higgsinos, and their mass eigenstates are neutralinos. 
So that, from (\ref{sHL}) and (\ref{sg}), neutralinos mediate two-body decays
$\tilde{G}\rightarrow \gamma\nu, Z\nu, V_{R}\nu$. 
In particular $\tilde{G}\rightarrow \gamma \nu$ is the 
particularly interesting since it 
may be constrained by 
very high energy gamma rays and neutrinos.
A peaked 2-body decay distribution  
is predicted.
The associated decay rate is 
\be{Gamma}
\Gamma^{(0)}_{\tilde{G}\rightarrow \gamma \nu}= \frac{\mu_{0}}{32\pi}\cos^{2}\theta_{W}\frac{m_{\nu}}{m_{\chi}}\frac{m_{\tilde{G}}^{3}}{M_{Pl}^{2}}\left(1-\frac{m_{\nu}^{2}}{m_{\tilde{G}}^{2}} \right)^{3}\left(1+\frac{m_{\nu}^{2}}{3 m_{\tilde{G}}^{2}} \right)
\ee
where $m_{\nu}$ is taken equal to 
 heaviest neutrino, 
assumed to be $m_{\nu_{3}}\simeq 0.07\, {\rm eV}$;
$$\mu_{0} \sim \frac{(\langle \phi_{S} \rangle)^{n-1}}{M^{n-1}}\,.$$
Let us note that in the case of $n=1$, the gravitino is rapidly destabilized
and the model should be easily ruled out 
In high scale supersymmetry breaking, 
assuming 
$m_{\chi}\simeq 10^{13}\, \rm GeV$
and $m_{\tilde{G}}\simeq 10^{11}\, \rm GeV$, 
the decay rate is of only 
$\Gamma^{0}\simeq 10^{-20}\, \rm eV$
corresponding to $\tau^{0} \simeq 10^{5}\, \rm s$. 
For a cosmologically stable gravitino
the decay rate must be suppressed down to $1\, \rm Gyr$ or so, 
i.e. of $10^{-11\div -12}$ orders. 

It should be noted that if gravitino lifetime is smaller than the age of the Universe, physics of the corresponding Unstable Dark Matter scenario should involve some additional stable particles - candidates to the modern dark matter. Moreover high energy neutrino and gamma background from gravitino decay lead to observable consequences \cite{sedelnikov,KhCh} that may exclude this possibility. 
If the R-symmetry is spontaneously broken before inflation,
$\langle \phi_{S}\rangle \simeq 10^{15}\, \rm GeV$,
Assuming $M\simeq M_{Pl}$, we must have 
$(\langle \phi_{s}\rangle/M_{Pl})^{(n-1)}\simeq 10^{-11\div 12}$. 
This may be obtained for $n=4$. 
Operators with $n<4$ are excluded, 
while operator with $n>4$ seems to much suppressed to be 
phenomenologically interesting. 

As a consequence, there is the interesting possibility of super-heavy gravitino decays 
$\tilde{G}\rightarrow \gamma \nu$ with two photons  and neutrino peaks 
of energy $E_{CM}\simeq m_{\tilde{G}}/2\simeq 10^{8}\div 10^{13}\, \rm GeV$. 
The observation of a so high energy neutrinos and photons could be a strong indirect evidence in 
favor of our scenario. In particular, these very high energy neutrinos can be observed by 
AUGER, Telescope Array, ANTARES and IceCube. 
and while eventually they could not be explained by any possible astrophysics sources.

\section{New challenges for string phenomenology }

The $R+\zeta R^{2}$ supergravity could be UV completed by string theory. 
Often in literature and in textbooks,
one can find the following statement: 
in the limit of $\alpha'=l_{s}^{2}\rightarrow 0$, 
 {\it superstrings reduce to supergravity models. }
 However, this is not completely corrected. 
 For instance, 
non-perturbative stringy corrections can 
generate new effective superpotential terms, 
even if not allowed by abelian symmetries at perturbative level.  
As a consequence, non-perturbative stringy corrections 
may destabilize the gravitino.
This may have dangerous or phenomenologically 
healthy implications (discussed above) for our model 
depending on the unknown global proprieties of 
the Calabi-Yau compactification. 

It is conceivable that the initial $U(1)_{R}$ gauge symmetry may
be broken
by Euclidean D-brane instantons 
 of open superstring theories 
or worldsheet instantons in heterotic superstring theory
(See \cite{Bianchi:2009ij} for a review on this subject). 

For example, a
$\mu HL$ superpotential 
can be generated by
 $E2$-branes 
 in intersecting D6-brane models 
 was discussed in Refs. \cite{Addazi:2015fua}
 (See also Refs.\cite{Addazi:2014ila,Addazi:2015ata,Addazi:2015rwa,Addazi:2015hka,Addazi:2015ewa,Addazi:2015yna,Addazi:2015eca,Addazi:2016mtn,Addazi:2015fua,Addazi:2015oba,Addazi:2015goa,Addazi:2016xuh}). 
The term $\mu HL$ is phenomenologically dangerous.
In fact, we must consider its interplaying with 
gravitino couplings with gauge bosons 
$W^{\pm},Z,\gamma,V_{R}$ and their related gauginos,
as mentioned above. As discussed in the previous section 
this should imply a very fast gravitino decay. 
So that, non-perturbative stringy instantons 
generating the $\mu HL$ superpotential 
must be suppressed in non-perturbative regime. 
If specific non-perturbative RR or NS-NS fluxes are wrapped by 
the instantonic Euclidean D-brane, such a suppression may be possible \cite{Bianchi:2012pn}.
Calling $\mathcal{N}_{N.P.}$ the non-pertubative suppression factor, 
this can screen the the bare decay rate as 
$\Gamma=\mathcal{N}_{N.P.}\Gamma_{0}$. 
A suppression factor $\mathcal{N}_{N.P.}\simeq 10^{-11}$
in order to get a gravitino cosmological life-time of at least $1\, \rm Gyr$ or so.

\section{Conclusions }

In this paper, we have reviewed 
Superheavy gravitino dark matter in 
Starobinsky supergravity 
with supersymmetry broken at high scales. 
We have reviewed how 
gravitinos may be non-thermally produced during inflationary slow-roll. 
As a consequence parameters of the inflaton potential and of gravitino dark matter are interconnected.  
This model provides a new peculiar prediction: Super-Heavy Gravitinos are produced with a continuos mass spectrum, 
following the inflaton field
\footnote{
The parameters space of gravitinos mass may change if a consistent amount 
of Primordial Black Holes   \cite{Khlopov:1985jw,Khlopov:2004tn,Khlopov:2008qy} were produced during the early Universe. 
We did not consider this other possible contribution.}.

We have commented about possible phenomenological implications
of our scenario. In particular, our model suggests possible 
two-body decays $\tilde{G}\rightarrow \gamma \nu$ 
producing 
very high energy peaks of neutrinos and photons, 
of $E_{CM}\simeq 10^{6}\div 10^{10}\,{\rm TeV}$. The detection of these very high energy neutrinos 
with a peak-like two-body decay distribution could provide a strong indirect hint in favor of our model. 

Finally, we have commented on possible open issues
regarding the UV completion of Starobinsky supergravity model 
in superstring theories. 
In particular,  $U_{R}(1)$ is not enough to protect the
gravitino by non-perturbative stringy instantons. 
The gravitino would be destabilized 
very fast if operators mixing the Higgsino with neutrino
were generated by non-perturbative solutions. 
This seems to be another problem toward a 
realistic UV embedding of our model in string theory 
in addition to the problem of string moduli stabilization during inflation.

\vspace{1cm} 

{\large \bf Acknowledgments} 
\vspace{4mm}

A would like to thank Fudan University of Shanghai and Hefei USTC -
and in particular Yifu Cai and Antonino Marcian\`o - for hospitality during the preparation of this 
paper. 
The work by MK was performed within the framework of the Center FRPP supported by MEPhI Academic Excellence Project (contract 02.03.21.0005, 27.08.2013).



\begin{thebibliography}{99}
  
  
 \bibitem{S1}
A. A. Starobinsky, Phys. Lett. B {\bf 91}, 99 (1980).


\bibitem{Ade:2015lrj}
  P.~A.~R.~Ade {\it et al.} [Planck Collaboration],
  arXiv:1502.02114 [astro-ph.CO].
  
  \bibitem{9}
  S. Cecotti, 
  Phys. Lett. B {\bf190}, 86 (1987).
  
  \bibitem{11}
   S. Cecotti, S. Ferrara, M. Porrati and S. Sabharwal, 
   Nucl. Phys. B {\bf 306}, 160 (1988).
  
  
  \bibitem{5}
  R. Kallosh and A. Linde, 
  JCAP {\bf 1306}, 027 (2013) [arXiv:1306.3211 [hep-th]].
  
  \bibitem{6}
  R. Kallosh and A. Linde, 
  JCAP {\bf 1306}, 028 (2013) [arXiv:1306.3214 [hep-th]].
  


\bibitem{18a}
S. V. Ketov and A. A. Starobinsky, 
Phys. Rev. D {\bf 83}, 063512 (2011) [arXiv:1011.0240 [hep-th]]. 

\bibitem{18b}
S. V. Ketov and A. A. Starobinsky, 
JCAP {\bf 1208}, 022 (2012) [arXiv:1203.0805 [hep-th]]. 

\bibitem{18c}
S. V. Ketov and S. Tsujikawa, 
Phys. Rev. D {\bf 86}, 023529 (2012) [arXiv:1205.2918 [hep-th]].

\bibitem{18d}
J.~Ellis, D.~V.~Nanopoulos and K.~A.~Olive,
  Phys.\ Rev.\ Lett.\  {\bf 111} (2013) 111301
   Erratum: [Phys.\ Rev.\ Lett.\  {\bf 111} (2013) no.12,  129902]
  doi:10.1103/PhysRevLett.111.129902, 10.1103/PhysRevLett.111.111301
  [arXiv:1305.1247 [hep-th]].

\bibitem{18e}
W.~Buchmuller, V.~Domcke and K.~Kamada,
  Phys.\ Lett.\ B {\bf 726}, 467 (2013)
  doi:10.1016/j.physletb.2013.08.042
  [arXiv:1306.3471 [hep-th]].
  
  \bibitem{21}
  J.~Ellis, D.~V.~Nanopoulos and K.~A.~Olive,
  JCAP {\bf 1310} (2013) 009
  doi:10.1088/1475-7516/2013/10/009
  [arXiv:1307.3537 [hep-th]].
  
  \bibitem{22}
 F.~Farakos, A.~Kehagias and A.~Riotto,
  Nucl.\ Phys.\ B {\bf 876} (2013) 187
  doi:10.1016/j.nuclphysb.2013.08.005
  [arXiv:1307.1137 [hep-th]].




\bibitem{Ferrara:2013rsa}
  S.~Ferrara, R.~Kallosh, A.~Linde and M.~Porrati,
  Phys.\ Rev.\ D {\bf 88} (2013) no.8,  085038
  doi:10.1103/PhysRevD.88.085038
  [arXiv:1307.7696 [hep-th]].

\bibitem{Ferrara:2013pla}
  S.~Ferrara, A.~Kehagias and M.~Porrati,
  Phys.\ Lett.\ B {\bf 727} (2013) 314
  doi:10.1016/j.physletb.2013.10.027
  [arXiv:1310.0399 [hep-th]].
  

\bibitem{Ferrara:2014cca}
  S.~Ferrara and M.~Porrati,
  Phys.\ Lett.\ B {\bf 737} (2014) 135
  doi:10.1016/j.physletb.2014.08.050
  [arXiv:1407.6164 [hep-th]].

\bibitem{Ferrara:2015ela}
  S.~Ferrara, A.~Kehagias and M.~Porrati,
  JHEP {\bf 1508} (2015) 001
  doi:10.1007/JHEP08(2015)001
  [arXiv:1506.01566 [hep-th]].
  
\bibitem{Ozkan:2014cua} 
  M.~Ozkan and Y.~Pang,
  Class.\ Quant.\ Grav.\  {\bf 31}, 205004 (2014)
  doi:10.1088/0264-9381/31/20/205004
  [arXiv:1402.5427 [hep-th]].
  
  
\bibitem{Ferrara:2014kva}
  S.~Ferrara, R.~Kallosh and A.~Linde,
  JHEP {\bf 1410} (2014) 143
  doi:10.1007/JHEP10(2014)143
  [arXiv:1408.4096 [hep-th]].
  
\bibitem{Antoniadis:2014oya}
  I.~Antoniadis, E.~Dudas, S.~Ferrara and A.~Sagnotti,
  Phys.\ Lett.\ B {\bf 733} (2014) 32
  doi:10.1016/j.physletb.2014.04.015
  [arXiv:1403.3269 [hep-th]].
  
\bibitem{Dudas:2015eha}
  E.~Dudas, S.~Ferrara, A.~Kehagias and A.~Sagnotti,
  JHEP {\bf 1509} (2015) 217
  doi:10.1007/JHEP09(2015)217
  [arXiv:1507.07842 [hep-th]].
  
\bibitem{Ferrara:2015gta}
  S.~Ferrara, M.~Porrati and A.~Sagnotti,
  Phys.\ Lett.\ B {\bf 749} (2015) 589
  doi:10.1016/j.physletb.2015.08.066
  [arXiv:1508.02939 [hep-th]].
  
\bibitem{Ferrara:2016buf}
  S.~Ferrara and A.~Sagnotti,
  Fortsch.\ Phys.\  {\bf 64} (2016) 371.
  doi:10.1002/prop.201500054
  



\bibitem{Addazi:2016bus}
  A.~Addazi and M.~Y.~Khlopov,
  arXiv:1612.06417 [gr-qc], accepted in PLB. 


\bibitem{Addazi:2017rkc}
  A.~Addazi and S.~V.~Ketov,
  arXiv:1701.02450 [hep-th].



\bibitem{Addazi:2014ila}
  A.~Addazi and M.~Bianchi,
  JHEP {\bf 1412} (2014) 089
  doi:10.1007/JHEP12(2014)089
  [arXiv:1407.2897 [hep-ph]].

\bibitem{Addazi:2015ata}
  A.~Addazi,
  JHEP {\bf 1504} (2015) 153
  doi:10.1007/JHEP04(2015)153
  [arXiv:1501.04660 [hep-ph]].

\bibitem{Addazi:2015rwa}
  A.~Addazi and M.~Bianchi,
  JHEP {\bf 1507} (2015) 144
  doi:10.1007/JHEP07(2015)144
  [arXiv:1502.01531 [hep-ph]].
  
\bibitem{Addazi:2015hka}
  A.~Addazi and M.~Bianchi,
  JHEP {\bf 1506} (2015) 012
  doi:10.1007/JHEP06(2015)012
  [arXiv:1502.08041 [hep-ph]].


\bibitem{Addazi:2015ewa}
  A.~Addazi,
  arXiv:1510.02911 [hep-ph], to appear in the proceeding of 14th MGM, 2015 Rome,  C15-07-12 .
  
  
\bibitem{Addazi:2015yna}
  A.~Addazi, M.~Bianchi and G.~Ricciardi,
  JHEP {\bf 1602} (2016) 035
  doi:10.1007/JHEP02(2016)035
  [arXiv:1510.00243 [hep-ph]].
  
\bibitem{Addazi:2015eca}
  A.~Addazi,
  Mod.\ Phys.\ Lett.\ A {\bf 31} (2016) no.17,  1650109
  doi:10.1142/S0217732316501091
  [arXiv:1504.06799 [hep-ph]].
  
\bibitem{Addazi:2016mtn}
  A.~Addazi and M.~Khlopov,
  Mod.\ Phys.\ Lett.\ A {\bf 31} (2016) no.19,  1650111
  doi:10.1142/S021773231650111X
  [arXiv:1604.07622 [hep-ph]].
  
\bibitem{Addazi:2015fua}
  A.~Addazi,
  EJTP {\bf 13}, No. 35 (2016) 39-56, arXiv:1505.00625 [hep-ph].
  
\bibitem{Addazi:2015oba}
  A.~Addazi,
  Int.\ J.\ Mod.\ Phys.\ A {\bf 31} (2016) no.16,  1650084
  doi:10.1142/S0217751X16500846
  [arXiv:1505.02080 [hep-ph]].
  
  
\bibitem{Addazi:2015goa}
  A.~Addazi,
  Phys.\ Lett.\ B {\bf 757} (2016) 462
  doi:10.1016/j.physletb.2016.04.018
  [arXiv:1506.06351 [hep-ph]].
  
\bibitem{Addazi:2016xuh}
  A.~Addazi, J.~W.~F.~Valle and C.~A.~Vaquera-Araujo,
  Phys.\ Lett.\ B {\bf 759} (2016) 471
  doi:10.1016/j.physletb.2016.06.015
  [arXiv:1604.02117 [hep-ph]].
  
   
  
 
\bibitem{Khlopov:1984pf}
  M.~Y.~Khlopov and A.~D.~Linde,
  Phys.\ Lett.\ B {\bf 138} (1984) 265.
  doi:10.1016/0370-2693(84)91656-3
\bibitem{Khlopov:1993ye}
       M.~Y.~Khlopov, Y.~L.~Levitan, E.~V.~Sedelnikov and I.~M.~Sobol,
  Phys.\ Atom.\ Nucl.\  {\bf 57}, 1393 (1994)
  [Yad.\ Fiz.\  {\bf 57}, 1466 (1994)].
\bibitem{Khlopov:2015oda}
  M.~Khlopov,
  Symmetry {\bf 7}, no. 2, 815 (2015)
  doi:10.3390/sym7020815
  [arXiv:1505.08077 [astro-ph.CO]].
  
\bibitem{Chung:1998zb}
  D.~J.~H.~Chung, E.~W.~Kolb and A.~Riotto,
  Phys.\ Rev.\ D {\bf 59} (1999) 023501
  doi:10.1103/PhysRevD.59.023501
  [hep-ph/9802238].
  
  \bibitem{Adiabatic1}
  T. S. Bunch, J. Phys. A {\bf 13}, 1297 (1980).
  
  \bibitem{Adiabatic2}
  L. Parker and S. A. Fulling, Phys. Rev. D {\bf 9}, 341 (1974).
    
    
  \bibitem{Adiabatic3}  
  L. Parker, Phys. Rev. {\bf 183}, 1057 (1969).
  
  \bibitem{20}
  N. D. Birrell and P. C. W. Davies, {\it Quantum Fields in Curved Space} (Cambridge University Press, Cambridge, 1982).
    
  
\bibitem{Rychkov:2007uq} V.S.Rychkov and A.Strumia
Phys.\ Rev.\ D {\bf 75} (2007) 075011
doi:10.1103/PhysRevD.75.075011 [hep-ph/0701104].



  \bibitem{Bianchi:2009ij}
  M.~Bianchi and M.~Samsonyan,
  Int.\ J.\ Mod.\ Phys.\ A {\bf 24} (2009) 5737
  doi:10.1142/S0217751X09048022
  [arXiv:0909.2173 [hep-th]].
  
  \bibitem{Bianchi:2012pn}
  M.~Bianchi, A.~Collinucci and L.~Martucci,
  Fortsch.\ Phys.\  {\bf 60} (2012) 914
  doi:10.1002/prop.201200030
  [arXiv:1202.5045 [hep-th]].
  
\bibitem{Bajc:2005zf}
  B.~Bajc, A.~Melfo, G.~Senjanovic and F.~Vissani,
  Phys.\ Rev.\ D {\bf 73} (2006) 055001
  doi:10.1103/PhysRevD.73.055001
  [hep-ph/0510139].
  
  
\bibitem{Bertolini:2009es}
  S.~Bertolini, L.~Di Luzio and M.~Malinsky,
  Phys.\ Rev.\ D {\bf 81} (2010) 035015
  doi:10.1103/PhysRevD.81.035015
  [arXiv:0912.1796 [hep-ph]].
  
  
\bibitem{Bianchi:2009bg}
  M.~Bianchi, F.~Fucito and J.~F.~Morales,
  JHEP {\bf 0908} (2009) 040
  doi:10.1088/1126-6708/2009/08/040
  [arXiv:0904.2156 [hep-th]].
  
  
  
  
\bibitem{Khlopov:1985jw}
  M.~Khlopov, B.~A.~Malomed and I.~B.~Zeldovich,
  Mon.\ Not.\ Roy.\ Astron.\ Soc.\  {\bf 215} (1985) 575.
\bibitem{sedelnikov}
E.V.Sedelnikov, M.Yu.Khlopov, Phys.Atom.Nucl.{\bf 59} (1996) 1000


\bibitem{KhCh}
M.Yu.Khlopov, V.M.Chechetkin, Sov.J.Part.Nucl. {\bf 18} (1987) 267
\bibitem{Khlopov:2004tn}
  M.~Y.~Khlopov, A.~Barrau and J.~Grain,
  Class.\ Quant.\ Grav.\  {\bf 23} (2006) 1875
    doi:10.1088/0264-9381/23/6/004 [astro-ph/0406621].

\bibitem{Khlopov:2008qy}
  M.~Y.~Khlopov,
  Res.\ Astron.\ Astrophys.\  {\bf 10} (2010) 495
  doi:10.1088/1674-4527/10/6/001 [arXiv:0801.0116 [astro-ph]].




\end{thebibliography}
\end{document}